\documentclass[twocolumn,showpacs,prl,preprintnumbers,amsmath,amssymb,superscriptaddress]{revtex4}

\usepackage{graphicx}% Include figure files
\usepackage{dcolumn}% Align table columns on decimal point
\usepackage{bm}% bold math
\usepackage{hyperref}
\usepackage{latexsym}
\usepackage{times}

\newcommand{\bin}[2]{
  \left(
  \begin{array}{@{}c@{}}
    #1  \\ #2
  \end{array}
  \right)
}

\begin{document}

\title{Modularity from Fluctuations in Random Graphs and Complex Networks}

\author{Roger Guimer\`a}

\affiliation{Department of Chemical and Biological Engineering, Northwestern University, Evanston, IL 60208, USA}

\author{Marta Sales-Pardo}

\affiliation{Department of Chemical and Biological Engineering, Northwestern University, Evanston, IL 60208, USA}

\author{Lu\'{\i}s A. Nunes Amaral}

\affiliation{Department of Chemical and Biological Engineering, Northwestern University, Evanston, IL 60208, USA}

%%%%%%%%%%%%% ABSTRACT %%%%%%%%%%%%%%%%%%%%%%%%%%%%%%%

\begin{abstract}
The mechanisms by which modularity emerges in complex networks are not
well understood but recent reports have suggested that modularity may
arise from evolutionary selection. We show that finding the modularity
of a network is analogous to finding the ground-state energy of a spin
system. Moreover, we demonstrate that, due to fluctuations, stochastic
network models give rise to modular networks. Specifically, we show
both numerically and analytically that random graphs and scale-free
networks have modularity. We argue that this fact must be taken into
consideration to define statistically-significant modularity in
complex networks.
\end{abstract}

\pacs{89.75.Hc, 05.50.+q, 02.60.Pn, 89.75.Fb}

\date{\today}

\maketitle

%%%%%%%%%%%%% INTRODUCTION %%%%%%%%%%%%%%%%%%%%%%%%%%%%%%%

Statistical, mathematical, and model-based analysis of complex
networks have recently uncovered interesting unifying patterns in
networks from seemingly unrelated disciplines
\cite{watts98,barabasi99,amaral00,albert02,dorogovtsev02}. In spite of
these advances, many properties of complex networks remain elusive, a
prominent one being modularity \cite{girvan02, newman03}. For example,
it is a matter of common experience that social networks have
communities of highly interconnected nodes that are poorly connected
to nodes in other communities. Such modular structures have been
reported not only in social networks
\cite{girvan02,guimera03,newman03}, but also in biochemical networks
\cite{hartwell99}, food webs \cite{pimm79}, and the Internet
\cite{eriksen03}. It is widely believed that the modular structure of
complex networks plays a critical role in their functionality
\cite{hartwell99}. There is therefore a clear need to develop
algorithms to identify modules accurately
\cite{girvan02,newman03,eriksen03,newman04,radicchi04}.

More fundamentally, the mechanisms by which modularity emerges in
complex networks are not well understood. In biological
networks---both biochemical and ecological---researchers have
suggested that modularity increases robustness, flexibility, and
stability \cite{hartwell99,pimm79}. Similarly, in engineered networks,
it has been suggested that modularity is effective to achieve
adaptability in rapidly changing environments \cite{alon03}. It may
therefore seem that {\it evolutionary pressures} make networks
modular, implying that any successful model of complex networks should
take into account external factors that enhance modularity. Recently,
however, Sol\'e and Fern\'andez have pointed out that models without
any external pressure are able to give rise to modular networks
\cite{sole03}.

%%%%%%%%%%% SUMMARY OF THE PAPER %%%%%%%%%%%%%%%

In this Letter, we show that Erd\"os-R\'enyi (ER) random graphs, in
which any pair of nodes is connected with probability $p$
\cite{bollobas01}, have a high modularity. We show numerically and
analytically that this high modularity is due to fluctuations in the
establishment of links, which are {\it magnified} by the large number
of ways in which a network can be partitioned into
modules. Furthermore, we show that one obtains similar results when
considering scale-free networks \cite{barabasi99}. We conclude by
discussing how these results should be taken into consideration to
define statistically significant modularity in complex networks.

\bigskip

%%%%%%%%%%% MEASURE OF MODULARITY AND HAMILTONIAN %%%%%%%

Following the first quantitative definition of modularity
\cite{newman03,newman04}, several groups have proposed heuristic
algorithms to detect modules in complex networks. For a given
partition of the nodes of a network into modules, the modularity
$\mathcal{M}$ of this partition is defined as \cite{newman03}
\begin{equation}
\mathcal{M}\equiv\sum_{s=1}^{r}\left[\frac{l_{s}}{L}-
\left(\frac{d_s}{2L}\right)^2\right]\,,
\label{e-modularity}
\end{equation}
where $r$ is the number of modules, $L$ is the number of links in the
network, $l_{s}$ is the number of links between nodes in module $s$,
and $d_s$ is the sum of the degrees of the nodes in module $s$. This
definition of modularity implies that $\mathcal{M}\le1$ and that
$\mathcal{M}=0$ for a {\it random partition} of the nodes
\cite{newman03}. We define the modularity $M$ of a network as the
largest modularity of all possible partitions of the network
$M=\max\{\mathcal{M}\}$.

The problem of finding the modularity of a network with $S$ nodes is
therefore analogous to the standard statistical mechanics problem of
finding the ground-state energy of the Hamiltonian
$\mathcal{H}=-L\mathcal{M}$. Specifically, one can map the network
into a spin system by defining the variables $s_i\in\{1,2,\dots,S\}$
as the module to which node $i$ belongs
%
%\footnote{In principle, there might be as many modules as nodes in the
%network.},
%
and the couplings $J_{ij}$ as being 1 if nodes $i$ and $j$ are
connected in the network and 0 otherwise. Then, from
Eq.~(\ref{e-modularity}), one can demonstrate that
\begin{eqnarray}
\mathcal{H}= & - & \sum_{i,j}^S \frac{J_{ij}}{2}\;\delta_{s_i s_j}\\
& + &\sum_{i,j,k,l}^S \left[\frac{J_{ij}J_{kl}}{16L}\left( \delta_{s_i
s_j s_k s_l}+2\delta_{s_i s_j s_k}+\delta_{s_i
s_k}\right)\right].\nonumber
\end{eqnarray}
This Hamiltonian corresponds to an $S$-state Potts model with both
ferromagnetic and anti-ferromagnetic terms, and two-, three-, and
four-spin interactions. Therefore, it seems difficult to apply methods
used in problems that are similar but formally simpler, like the graph
coloring problem \cite{mulet02}. Rather, we propose here a
heuristic estimation of the modularity for a number of interesting
graph models, namely low-dimensional regular lattices, ER random
graphs \cite{bollobas01} and scale-free networks \cite{barabasi99}.

%%%%%%%%%%%% MODULARITY FOR LOW-D LATTICES %%%%%%%%%%%%%%%%%%%%%

{\it Low-dimensional regular lattices---} Consider a one-dimensional
lattice with $S$ nodes, each one connected to its two neighbors
\footnote{In this case, there are no fluctuations involved in the
creation of the network. Rather, modularity arises because neighbors
of a node in a low-dimensional lattice are also neighbors of each
other, and neither the node nor its neighbors are linked to nodes that
are far away in the lattice. As an example, consider the cities and
towns in Europe and all the roads between them. This ``road network''
is two-dimensional and has modules that correspond roughly to the
countries. In each of this ``sub-modules'' people have different
customs, food preferences, languages or dialects, etc, that is,
communities exist. It is also worth noting that the hypothesis used in
the calculations are essentially the same that we use for random
graphs and scale-free networks, and that the modularity of
one-dimensional regular lattices turns out to be useful in certain
limits of ER random graphs and scale-free networks. }.
%
%If the lattice is very small, no modules can be identified. However,
%if we increase the size of the network the maximum modularity
%corresponds to splitting the network into two modules; if the chain
%continues to grow, the maximum modularity corresponds to dividing the
%nodes into three modules, and so on.
%
This case is particularly simple because the modules comprise only
contiguous nodes and, therefore, the number of between-module links
equals the number $r$ of modules. Assuming that all modules have
approximately the same size $n=S/r$, the modularity of a partition
with $r$ modules is
\begin{equation}
\mathcal{M}_{1D}(S;r)=\frac{S-r}{S}-\frac{1}{r}\,,
\end{equation}
where we have used the fact that the number $L$ of links is $L\approx
S$. Under these assumptions, the problem of finding the modularity of
a regular one-dimensional lattice is reduced to finding the optimal
number $r^*$ of modules, that is, the number of modules that yields
the maximum modularity. One can show that $r^*(S)=\sqrt{S}$, and the
modularity is
\begin{equation}
M_{1D}(S)=1-\frac{2}{\sqrt{S}}\,.
\label{e-modlowd1}
\end{equation}
Note that the only assumption in the calculation is that all modules
have approximately the same number of nodes. Numerical results confirm
that this is a sensible assumption.

One can generalize this result to one-dimensional lattices in which
each node is connected to $z$ nodes on the left and $z$ on the
right. In this case, the leading contributions to the modularity are
\begin{equation}
M_{1D}(S,z)=1-\sqrt{\frac{2(z+1)}{S}}\;.
\label{e-modlowd2}
\end{equation}
Similarly, one can calculate the modularity of $d$-dimensional cubic
lattices in which each node is connected to $2z$ nodes in each one of
the $d$ directions, to obtain that
\cite{guimera??c}
\begin{equation}
M_{dD}(S,z)=1-(d+1)\left(\frac{z+1}{2d}\right)^{\frac{d}{d+1}}
\frac{1}{S^{\frac{1}{d+1}}}\;.
\label{e-modlowd3}
\end{equation}
%

%%%%%%%%%%%% CALCULATION OF THE MODULARITY FOR ER %%%%%%%%%%%%%%%

{\it Random graphs---} In ER random graphs \cite{bollobas01}, each
pair of nodes is connected with probability $p$. As for
$d$-dimensional lattices, we assume that the partition of the network
with highest modularity consists of $r$ modules with approximately the
same number of nodes $n=S/r$, the same number of within-module links
$k_i$, and the same number of links $k_o$ to other modules. In the
$S\gg1$ limit, we can assume that the total number of links is
$S^2p/2$ and, therefore, $k_i$ and $k_o$ are related by
\begin{equation}
k_o(S, p;r, k_i)=\frac{S^2p}{r}-2k_i.
\end{equation}
Hence, for $S\gg1$, the modularity of such a partition is simply
\begin{equation}
\mathcal{M}_{ER}(S, p;r, k_i)=\frac{2rk_i}{S^2p}-\frac{1}{r}\, .
\label{e-modr}
\end{equation}

Under these assumptions, the problem of finding the modularity of a
random graph is reduced to finding a partition of the graph with the
following properties: (i) The partition consists of $r$ equal modules,
each one with $k_i$ within-module links; (ii) The partition {\it
typically exists} in a random graph; and (iii) The partition yields
the maximum modularity relative to the other partitions that typically
exist.

In a random graph with $S$ nodes and linking probability $p$, the
average number $\mathcal{N}$ of different partitions with $r$
identical modules, each with $k_i$ links, is $\mathcal{N}(S, p;r,
k_i)$. A certain partition {\it typically exists} if $\mathcal{N}(S,
p;r, k_i)\ge1$. Among all the partitions that typically exist, we are
interested in the one whose modularity is maximum. In other words,
given a certain number $r$ of modules, we want a partition with as
many within-module links as possible. Therefore, if one finds a very
{\it common} partition $\mathcal{N}(S, p;r, k_i)\gg1$, it must be
possible to find another partition with the same $r$ and $k_i'>k_i$
that has larger modularity. This new partition will be {\it rarer}
than the former one $\mathcal{N}(S, p;r, k_i')<\mathcal{N}(S, p;r,
k_i)$. By iterating this argument, one concludes that the partition we
are interested in must satisfy
\begin{equation}
\mathcal{N}(S, p;r, k_i^*(S, p;r))=1\, , 
\label{e-condit}
\end{equation}
where $k_i^*(S, p;r)$ is the maximum number of within-module links
that one can typically find in a partition with $r$ identical modules.

To calculate $\mathcal{N}(S, p;r, k_i)$, we use the following
process. First, we calculate the number $\mathcal{N}_1$ of ways in
which a module of size $n=S/r$, with $k_i$ within-module links and
$k_o(r, k_i)$ external links, can be {\it separated} from the rest of
the graph:
\begin{equation}
\mathcal{N}_1=\bin{S}{n}P_i(S, p;n, k_i)P_o(S, p;n, k_o)\, , 
\end{equation}
where
\begin{eqnarray}
P_i(S, p;n, x) & = &
\bin{\frac{n(n-1)}{2}}{x}p^{x}(1-p)^{\frac{n(n-1)}{2}-x}\,,
\label{e-pi}\\
P_o(S, p;n, x) & = & \bin{n(S-n)}{x}p^{x}(1-p)^{n(S-n)-x}\,
. \label{e-po}
\end{eqnarray}
The next step is to separate the second module from the remaining set
of $S-n$ nodes. It is important to note that the second module only
needs to establish $k_o (1-n/(S-n))$ external links, because the
remaining $k_o n/(S-n)$ are already established with the first
module. Therefore,
\begin{equation}
\mathcal{N}_2=\bin{S-n}{n}P_i(S, p;n, k_i)P_o(S, p;n,
k_o(1-\frac{n}{S-n}))\, .
\end{equation}
Repeating this separation process, one can see that the general term
is of the form
\begin{equation}
\mathcal{N}_{t+1}=\bin{S-tn}{n}P_i(S, p;n, k_i)P_o(S, p;n, k_o(1-\frac{tn}{S-n}))\, .
\label{e-Nk}
\end{equation}
Finally, $\mathcal{N}(S, p;r, k_i)$ is the product of all the
individual module separations
\begin{equation}
\mathcal{N}(S, p;r, k_i, k_o(r,k_i))=\prod_{t=1}^{r}
\mathcal{N}_t\,,
\label{e-finalN}
\end{equation}
so that Eq.~(\ref{e-condit}) can be solved numerically to obtain
$k_i^*(S, p;r)$ using Eqs.~(\ref{e-pi}), (\ref{e-po}), (\ref{e-Nk}),
and (\ref{e-finalN}).

Once we find $k_i^*(S, p;r)$ for a given value of $r$, we use
Eq.~(\ref{e-modr}) to obtain the modularity. Finally, we select the
optimal number of modules $r=r^*(S, p)$ and the modularity $M_{ER}(S,
p)$ of the ER random graph is
\begin{equation}
M_{ER}(S, p)=\frac{2r^*(S,p)k_i^*(S,p;r^*)}{S^2p}-\frac{1}{r^*(S,p)}\,.
\label{e-modfinal}
\end{equation}

\begin{figure}[t!] 
  \includegraphics*[width=0.53\columnwidth]{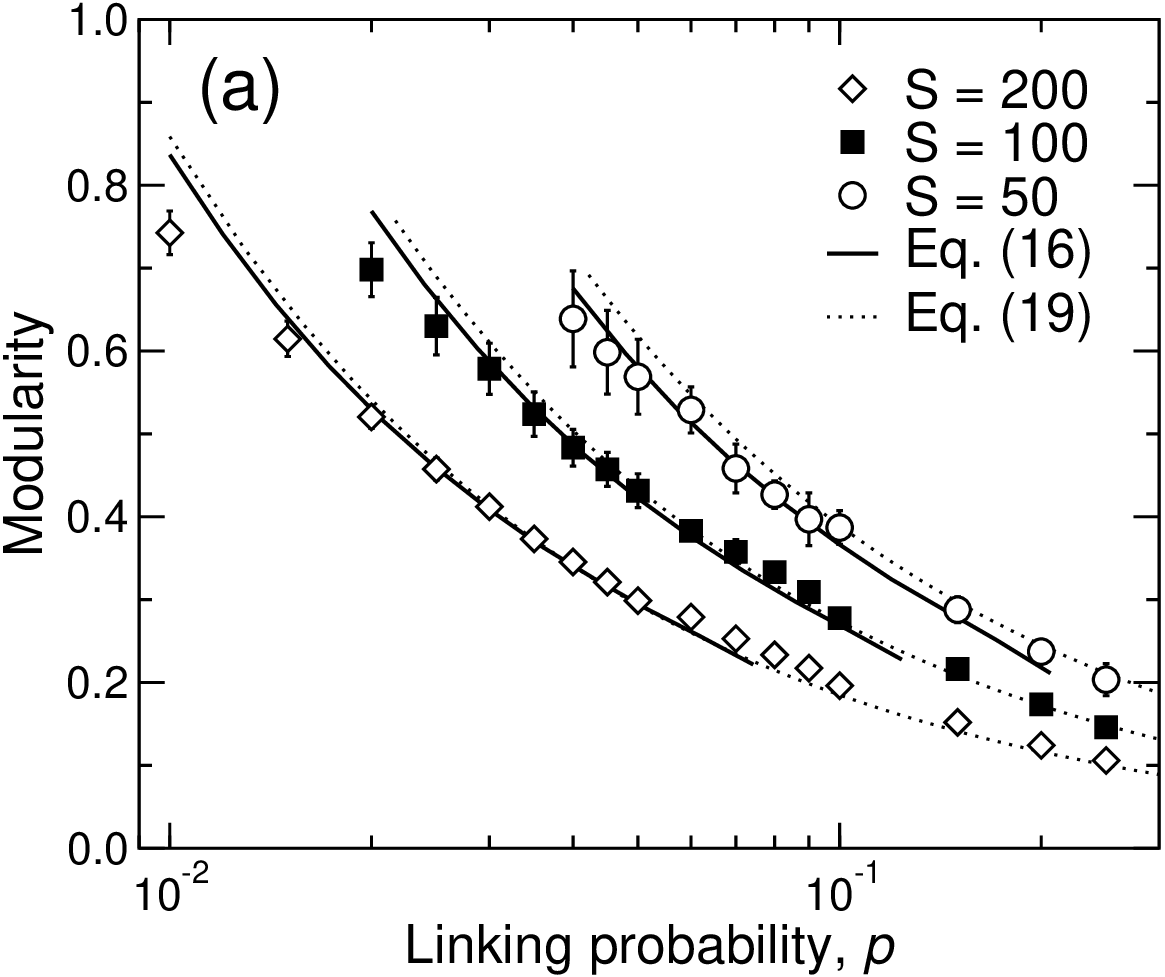}
  \includegraphics*[width=0.53\columnwidth]{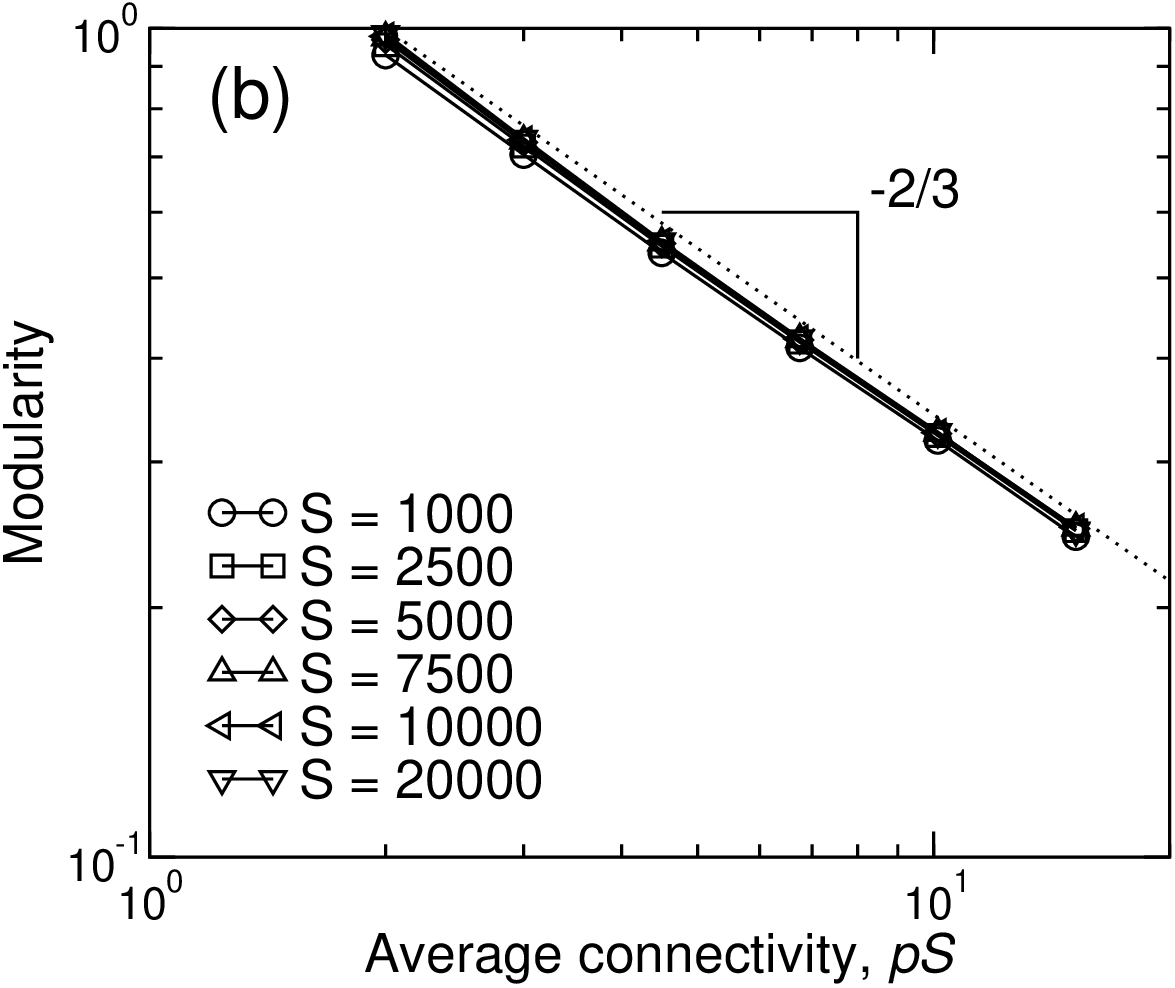}
  \caption{Modularity in Erd\"os-R\'enyi random graphs.
  (a) Comparison of numerical results of the modularity as a function
  of the linking probability, and the predictions of
  Eqs.~(\ref{e-modfinal}) and (\ref{e-modERfinal}). The numerical
  results are obtained by maximizing the modularity,
  Eq.~(\ref{e-modularity}), using simulated annealing
  \cite{kirkpatrick83}.
  (b) Modularity as a function of $pS$ for large networks, as
  predicted by Eq.~(\ref{e-modfinal}). Both in (a) and (b), numerical
  problems in the solution of Eq.~(\ref{e-condit}) prevent us from
  obtaining values of the modularity for larger values of $p$.
  }
  \label{f-er} 
\end{figure}
In Fig.~\ref{f-er}(a), we compare the modularity of ER graphs
obtained through optimization of Eq.~(\ref{e-modularity}) using
simulated annealing \cite{kirkpatrick83}, with the predictions of
Eq.~(\ref{e-modfinal}). We find good agreement in the relevant region
of sparse but connected graphs, that is, $2/S<p\ll 1$.

Equation ~(\ref{e-modfinal}) enables us to obtain the modularity of
large random graphs, something that would not be possible using
simulated annealing because of the computational cost. In
Fig.~\ref{f-er}(b) we show that for $S\rightarrow\infty$ the
modularity only depends on $pS$
\begin{equation}
M_{ER}(S\rightarrow\infty, p)\sim (pS)^{-2/3}.
\label{e-modERS}
\end{equation}
To obtain a closed expression for $M_{ER}$ for any value of $S$, we
note that at the percolation point $pS=2$ the random graph contains
essentially no loops, that is, the graph is a tree
\cite{bollobas01}. In this case, one can find partitions in which the
number of between-module links equals the number of modules $r$ as in
the simple one-dimensional case, and the modularity is
\begin{equation}
M_{ER}(S,p=2/S)=M_{1D}(S)=1-\frac{2}{\sqrt{S}}\,.
\label{e-modERcrit}
\end{equation}

We propose the simplest {\it ansatz} that verifies
Eqs.~(\ref{e-modERS}) and (\ref{e-modERcrit}) simultaneously
\begin{equation}
M_{ER}(S,p)=\left(1-\frac{2}{\sqrt{S}}\right)
\left(\frac{2}{pS}\right)^{2/3}\,.
\label{e-modERfinal}
\end{equation}
In Fig.~\ref{f-er}(a), we show that Eq.~(\ref{e-modERfinal}) is in
good agreement with values obtained using simulated annealing.

Our analytic treatment allows us to explain the origin of the
modularity in random graphs. The typical partition of an ER graph into
modules of size $n$ is very unlikely to have a number of within-module
links $k_i$ larger than the average $pn(n-1)/2$, expected for a random
partition of the nodes. However, the number of possible partitions
$S!/(n!r)$ is so large that, typically, there exists a partition whose
$k_i$ is much larger than the average. For example, for a network with
$S=200$ and $p=0.02$ one typically finds a partition with $r=7$
modules and $k_i\approx36$, instead of the value $k_i\approx8$
expected for a random partition.

Remarkably, the modularity of a random graph can be as large as that
of a graph with modular structure imposed at the onset
\cite{girvan02}. In such a graph, nodes are divided into modules and
each pair of nodes is connected with probability $p_i$ if they belong
to the same module, and with probability $p_o<p_i$ otherwise. Using
the same example as before, the modularity of an ER graph with $S=200$
and $p=0.02$ is the same as the modularity of a graph with $m=7$
modules, $p_i\approx0.09$, and $p_o\approx0.004$.

\bigskip

%%%%%%%%%%%% CALCULATION OF THE MODULARITY FOR SF %%%%%%%%%%%%%%%

{\it Scale-free networks---}So far, we have considered $d$-dimensional
regular lattices and ER random graphs, in which all nodes have
essentially the same degree. However, many complex networks display
scale-free degree distributions \cite{albert02}, meaning that some
nodes have degrees that are orders of magnitude larger than the
average. Since the results presented for ER graphs rely on the fact
that there are many partitions of the network and implicitly on the
fact that nodes are exchangeable, it is worth asking whether
``random'' scale-free networks also display modularity.

\begin{figure}[tb] 
  \includegraphics*[width=0.53\columnwidth]{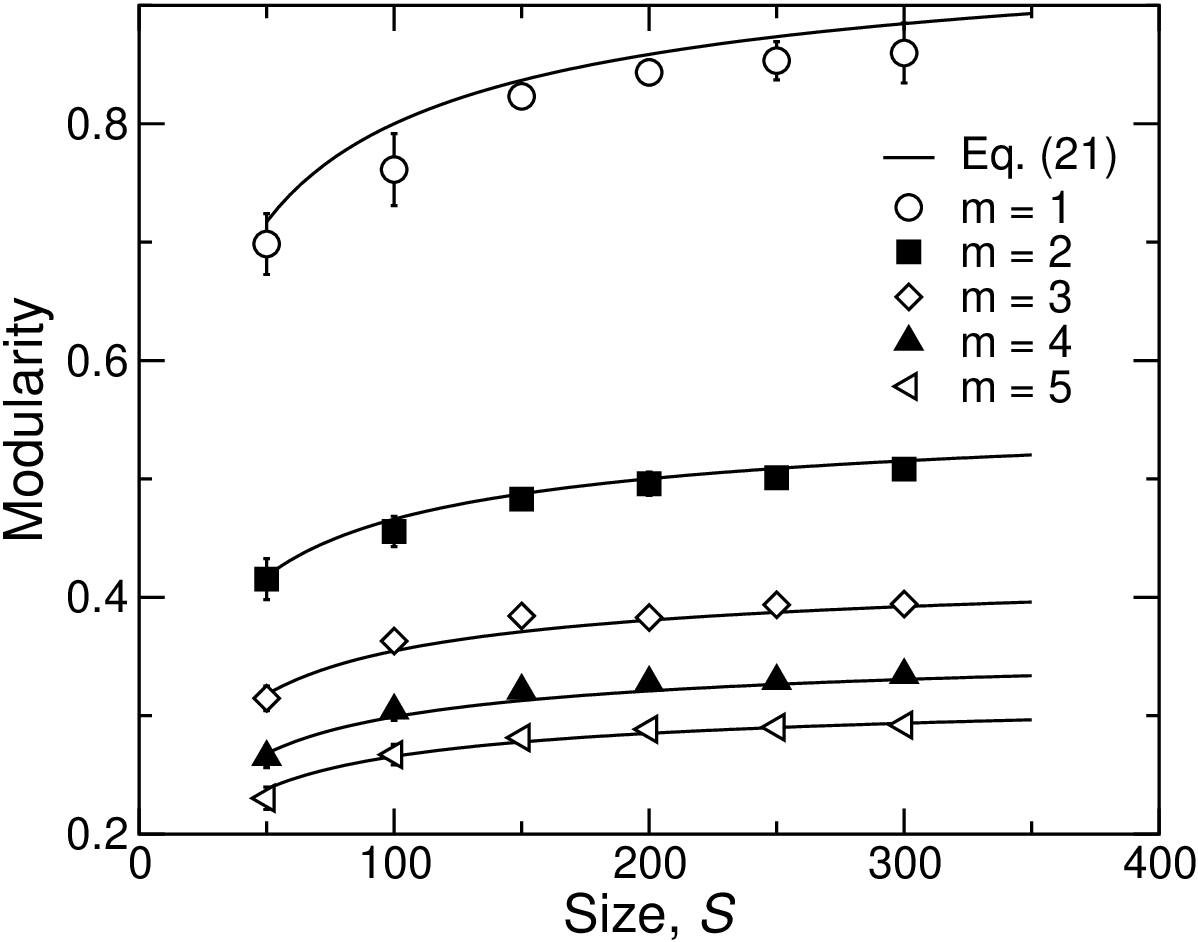}
  \caption{Modularity in scale-free networks. Numerical results of the
  modularity as a function of the network size $S$ for different
  values of $m$. These results are obtained by maximizing the
  modularity, Eq.~(\ref{e-modularity}), with simulated annealing. The
  lines are the predictions of Eq.~(\ref{e-modba}), with
  $a=0.165\pm0.009$ in all the cases.
  }
  \label{f-ba} 
\end{figure}

To answer this question, we use the scale-free model proposed in
\cite{barabasi99}. In the model, the network grows by the addition of
new nodes. Each time a new node is added, it establishes $m$
preferential connections to nodes already in the network. In
Fig.~\ref{f-ba}(a), we show the modularity of scale-free networks as a
function of the network size $S$ for different values of $m$. As
before, we find the modularity by optimizing Eq.~(\ref{e-modularity})
using simulated annealing. As for ER graphs, the modularity approaches
a finite value for large $S$ and decreases with the connectivity $m$.

We are unable to derive a general expression for the modularity of
scale-free networks. However, for $m=1$ the scale-free network is a
tree. Thus,
\begin{equation}
M_{SF}(S,m=1)=M_{1D}(S)=1-\frac{2}{\sqrt{S}}\,.
\label{e-modbam1}
\end{equation}

For larger values of $m$, we find numerically that, at a fixed network
size, the modularity is a linear function of $1/m$. The simplest
possible {\it ansatz} for the modularity that verifies this condition
and Eq.~(\ref{e-modbam1}) simultaneously is
\begin{equation}
M_{SF}(S,m)=\left(a+\frac{1-a}{m}\right)\left(1-\frac{2}{\sqrt{S}}
\right)\,.
\label{e-modba}
\end{equation}
As we show in Fig.\ref{f-ba}, this approximation works well for
$a=0.165\pm0.009$.
%
%
%

%%%%%%%%%%%% CONCLUSIONS %%%%%%%%%%%%%%%

{\it Conclusions---} We have shown that modularity in networks can
arise due to a number of mechanisms. We have demonstrated that
networks embedded in low dimensional spaces have high modularity. We
have also shown analytically and numerically that, surprisingly,
random graphs and scale-free networks have high modularity due to
fluctuations in the establishment of links.

Recently, several works have reported the existence of modules in
complex networks and suggested that some evolutionary mechanism must
enhance modularity. This statement is based, in the best of the cases,
on the fact that the modularity is {\it large enough}, and relies
implicitly on the assumption that random graphs have low modularity.

Our results enable one to define {\it statistically significant}
modularity in networks. We argue that, just as it is already done for
the clustering coefficient and other quantities, the modularity of
complex networks must always be compared to the {\it null case} of a
random graph. The analytical expressions we have derived provide a
convenient way to carry out such a comparison.

%%%%%%%%%%%% ACKNOWLEDGMENTS %%%%%%%%%%%%%%%

\begin{acknowledgments}
We thank Alex Arenas, Andr\'e A. Moreira, Carla A. Ng, and Daniel
B. Stouffer for numerous suggestions and discussions. R.G. and
M.S. thank the Fulbright Program and the Spanish Ministry of
Education, Culture \& Sports. L.A.N.A. gratefully acknowledges the
support of a Searle Leadership Fund Award and of a NIH/NIGMS K-25
award.
\end{acknowledgments}

%%%%%%%%%%%% REFERENCES %%%%%%%%%%%%%%%

%\bibliography{/home/projects/BibTeX/strings,/home/projects/BibTeX/roger}

\end{document}